\documentclass{emulateapj}

\usepackage{natbib}
\usepackage{graphicx}

\bibpunct{(}{)}{;}{a}{}{,}
\begin{document}                          

\title{Control of Star Formation in Galaxies by Gravitational Instability}    

\author{Yuexing Li, Mordecai-Mark Mac Low}
\affil{Department of Astronomy, Columbia University, New York,
NY 10027, USA}
\affil{Department of Astrophysics, American Museum of Natural
History, New York, NY 10024, USA}
\and 
\author{Ralf S. Klessen}
\affil{Astrophysikalisches Institut Potsdam, An der Sternwarte
16, D-14482 Potsdam, Germany} 
\email{yxli@astro.columbia.edu, mordecai@amnh.org, rklessen@aip.de} 

\begin{abstract}
We study gravitational instability and consequent star formation in a wide
range of isolated disk galaxies, using three-dimensional, smoothed particle
hydrodynamics simulations at resolution sufficient to fully resolve
gravitational collapse. Stellar feedback is represented by an isothermal
equation of state. Absorbing sink particles are inserted in dynamically bound,
converging regions with number density $n > 10^3$~cm$^{-3}$ to directly
measure the mass of gravitationally collapsing gas available for star formation. 
Our models quantitatively reproduce not only the observed Schmidt law, but
also the observed star formation threshold in disk galaxies.  Our results
suggest that the dominant physical mechanism determining the star formation
rate is just the strength of gravitational instability, with feedback
primarily functioning to maintain a roughly constant effective sound speed.
\end{abstract}
 

\section{INTRODUCTION}
Stars form in galaxies at hugely varying rates \citep{kennicutt98a}.
The mechanisms that control the star formation rate from interstellar
gas are widely debated \citep*{shu87, elmegreen02, larson03, mk04}.
Gravitational collapse is opposed by gas pressure, supersonic
turbulence, magnetic fields, and rotational shear.  Gas pressure in
turn is regulated by radiative cooling and stellar and turbulent
heating. Despite this complexity, star-forming spiral galaxies follow
two empirical laws. First, stars only form above a critical
gas surface density \citep{martin01} that appears to be determined by the
\citet{toomre64} criterion for gravitational instability. 
Second, the rate of star formation is proportional
to a power of the total gas surface density \citep{schmidt59, kennicutt98b}.

A number of groups have simulated disk galaxies in isolation or in mergers, or
in cosmological contexts, e.g., \citet*{mihos94, friedli95, sommer-larsen99,
springel00, barnes02, governato04}. \citet{robertson04} review
this work. However, in these simulations, star formation is generally set up with
empirical recipes \textit{a priori}. The origin of the observed Schmidt
law remains unclear. 

Recent cosmological simulations with moderate mass resolution by
\citet{kravtsov03} show that the Schmidt law is a manifestation of the
overall density distribution of the ISM, and find little contribution from
feedback. However, the strength of gravitational instability was not directly
measured in his work, so a direct connection could not be made between
instability and the Schmidt Law, as we do here. The importance of
gravitational instability in controlling large-scale star formation was
emphasized by \citet{friedli95} and \citet{elmegreen02}. A similar
conclusion comes from the observation that thin dust lanes in galaxies
only form in gravitationally unstable regions \citep*{dalcanton04}.

We simulate a large set of isolated galaxies to investigate
gravitational instability and consequent star formation.  In this
Letter, we examine star formation as a function of gravitational
instability, and compare the global Schmidt law and star formation
thresholds derived from our simulations to the observations.

\section{COMPUTATIONAL METHOD}
\label{sec_com}
We use the smoothed particle hydrodynamics (SPH) code GADGET
\citep*{springel01}, modified to include absorbing sink particles
\citep*{bate95} to directly measure the mass of gravitationally
collapsing gas. Sink particles, representing star clusters
(SCs), replace gravitationally bound regions of converging flow that reach 
number densities $n > 10^3$~cm$^{-3}$. (These regions have pressures $P/k \sim 
10^7$~K~cm$^{-3}$ typical of star-forming regions.)

Our galaxy model consists of a dark matter halo, and a disk of stars
and isothermal gas.  The galaxy structure is based on the analytical
work by \citet*{mo98}, as implemented numerically by \citet{springel99}
and \citet{springel00}. The isothermal sound speed is chosen to be
either $c_1 = 6$~km~s$^{-1}$ in models with low temperature $T$ or
$c_2 = 15$~km~s$^{-1}$ in high $T$ models.  Table 1 lists the most
important model parameters. The Toomre criterion for gravitational
instability that couples stars and gas, $Q_{\rm sg}$ is calculated
following \citet{rafikov01}, and the minimum value is derived using the
wavenumber $k$ of greatest instability and lowest $Q_{sg}$ at each radius.

The gas, halo and disk particles are distributed with number ratio
$N_{\rm g}$ : $N_{\rm h}$ : $N_{\rm d}$ = 5 : 3 : 2. The gravitational
softening lengths of the halo $\epsilon_{\rm h} = 0.4$~kpc and disk
$\epsilon_{\rm d}= 0.1$~kpc, while that of the gas
$\epsilon_{\rm g}$ is given in Table~1 for each model.  
The minimum spatial and mass resolutions in the gas are given by 
$\epsilon_{\rm g}$ and twice the kernel mass ($\sim
80 m_g$).  We adopt typical values for the halo concentration
parameter $c = 5$, spin parameter $\lambda = 0.05$, and Hubble
constant $H_0 = 70$ km s$^{-1}$ Mpc$^{-1}$ \citep{springel00}.  The
spin parameter used is a typical one for galaxies subject to the tidal
forces of the cosmological background.
\citet{reed03} suggest a wide range of $c$ for galaxy-size
halos. However, this parameter is based on a simple model of the halo
formation time \citep*{nfw97}, with poorly known distribution
\citep{mo98}. \citet{springel99} suggest that $c = 5$ is theoretically
expected for flat, low-density universes. 

\begin{deluxetable}{lllllll}
\tablecolumns{7}
\tablecaption{Galaxy Models and Numerical Parameters \label{tab1}
}
\tablehead{\colhead{Model\tablenotemark{a}} & 
\colhead{$f_{\rm g}$\tablenotemark{b}} & 
\colhead{$Q_{\rm sg}$(LT)\tablenotemark{c}} & \colhead{$Q_{\rm
sg}$(HT)\tablenotemark{d}} & \colhead{$N_{\rm tot}$\tablenotemark{e}} & 
\colhead{$\epsilon_{\rm g}$\tablenotemark{f}} & \colhead{$m_{\rm
g}$\tablenotemark{g}}}  
\startdata
G50-1  & 1   &  1.22  &  1.45  & $1.0 $ & 10 & $ 0.08 $\\
G50-2  & 2.5 &  0.94  &  1.53  & $1.0 $ & 10 & $ 0.21 $\\
G50-3  & 4.5 &  0.65  &  1.52  & $1.0 $ & 10 & $ 0.37 $\\
G50-4  & 9   &  0.33  &  0.82  & $1.0 $ & 10 & $ 0.75 $\\
G100-1 &  1   &  1.08  &  \nodata  & $6.4 $ & 7 & $ 0.10 $\\
G100-1 &  1   &  \nodata  &  1.27  & $1.0 $ & 10 & $ 0.66 $\\
G100-2 &  2.5 &  \nodata  &  1.07  & $1.0 $ & 10 & $ 1.65 $\\
G100-3 &  4.5 &  \nodata  &  0.82  & $1.0 $ & 10 & $ 2.97 $\\
G100-4 &  9   &  \nodata  &  0.42  & $1.0 $ & 20 & $ 5.94 $\\
G120-3 &  4.5 &  \nodata  &  0.68  & $1.0 $ & 20 & $ 5.17 $\\
G120-4 &  9   &  \nodata  &  0.35  & $1.0 $ & 30 & $ 10.3 $\\
G160-1 &  1   &  \nodata  &  1.34  & $1.0 $ & 20 & $ 2.72 $\\
G160-2 &  2.5 &  \nodata  &  0.89  & $1.0 $ & 20 & $ 6.80 $\\
G160-3 &  4.5 &  \nodata  &  0.52  & $1.0 $ & 30 & $ 12.2 $\\ 
G160-4 &  9   &  \nodata  &  0.26  & $1.5 $ & 40 & $ 16.3 $\\
G220-1 &  1   &  0.65  & \nodata   & $6.4 $ & 15 & $ 1.11 $\\
G220-1 &  1   &  \nodata  &  1.11  & $1.0 $ & 20 & $ 7.07 $\\
G220-2 &  2.5 &  \nodata  &  0.66  & $1.2 $ & 30 & $ 14.8 $\\
G220-3 &  4.5 &  \nodata  &  0.38  & $2.0 $ & 40 & $ 15.9 $\\
G220-4 &  9   &  \nodata  &  0.19  & $4.0 $ & 40 & $ 16.0 $\\          
\enddata         
\tablenotetext{a}{First number is rotational velocity 
in km s$^{-1}$ at virial radius.} 
\tablenotetext{b}{Percentage of total halo mass in gas.}
\tablenotetext{c}{Minimum initial $Q_{\rm sg}$ for low $T$ model.
  Missing data indicates models not run at full resolution.}
\tablenotetext{d}{Minimum initial $Q_{\rm sg}$ for high $T$ model}
\tablenotetext{e}{Millions of particles in high resolution runs.}
\tablenotetext{f}{Gravitational softening length of gas in pc.}
\tablenotetext{g}{Gas particle mass in units of $10^4 M_{\odot}$.}      
\end{deluxetable}

Models of gravitational collapse must satisfy three numerical criteria: 
the Jeans resolution criterion (\citealt{bate97}, hereafter BB97;
\citealt{whitworth98}), the gravity-hydro balance criterion for gravitational
softening (BB97), and the equipartition criterion for particle masses
\citep{steinmetz97}. \citet{truelove98} suggest that a Jeans mass must be
resolved with far more than the $N_k = 2$ smoothing kernels proposed by
BB97. Therefore we performed a resolution study of model G100-1 (LT) with
$N_{\rm tot}$ = $10^5$, $8\times 10^5$, and $6.4 \times 10^6$, corresponding
to $N_{\rm k}\approx$ 0.4, 3.0 and 23.9, respectively. We find convergence to
within 10\% of the global amount of mass accreted by sink particles between
the two highest resolutions, suggesting that the BB97 criterion is sufficient
for the problem considered here. 

We performed 24 simulations satisfying all three criteria, including six
models of low mass galaxies with low $T$ to study the effect of changing the
effective sound speed. We also set a minimum value of $N_{\rm tot}
\ge 10^6$ particles for lower mass galaxies resolved with fewer particles. 

\section{GLOBAL SCHMIDT LAW}

To derive the Schmidt law, we average $\Sigma_{\rm SFR}$ and
$\Sigma_{\rm gas}$ over the star forming region following
\citet{kennicutt89}, with radius chosen to encircle 80\% of the mass
in sinks. To estimate the star formation rate, we make the assumption
that individual sinks represent dense molecular clouds that form stars at some
efficiency. Observations by \citet{rownd99} suggest that the {\em local} star
formation efficiency (SFE) in molecular clouds remains roughly
constant. \citet{kennicutt98b} shows a median SFE of 30\% in starburst
galaxies dominated by molecular gas. This suggests the local SFE of dense
molecular clouds around 30\%.  We therefore adopt a fixed local SFE of
$\epsilon$ = 30\% to convert the mass of sinks to stars. Note that this local
efficiency is different from the global star formation efficiency in galaxies,
which measures the fraction of the {\em total} gas turned into stars. The
global SFE can range from 1--100\% \citep{kennicutt98b}, depending
on the gas distribution and the molecular gas fraction. 

\begin{figure}[h]
\begin{center}
\includegraphics[width=2.7in]{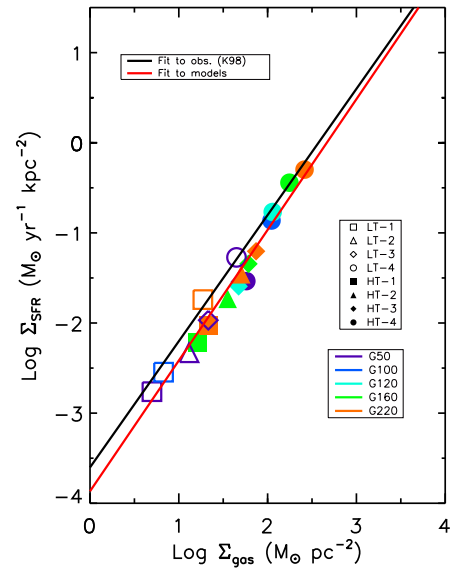}
\caption{\label{fig_global} 
Schmidt law from fully resolved low ({\em open symbols}) and high ({\em filled
symbols}) $T$ models listed in Table~\ref{tab1} that showed gravitational collapse.
The colors indicate the galaxy rotational velocities, while the symbol
shapes indicate the gas fractions, as specified in the legend.  The
black line is the best fit to the observations from
\citet{kennicutt98b}, while the red line is the best fit to the
simulations.}
\end{center} 
\end{figure} 

Figure \ref{fig_global} shows the Schmidt law derived from our simulations.
The best fit to the observations by \citet{kennicutt98b} gives a
Schmidt law $\Sigma_{\rm {SFR}} = A \Sigma_{\rm gas}^{\alpha}$ with
global efficiency $A = (2.5 \pm 0.7) \times 10^{-4}$ and power law
$\alpha = 1.4 \pm 0.15$, where $\Sigma_{\rm {SFR}}$ is given in units
of $M_{\odot}$~kpc$^{-2}$~yr$^{-1}$, and $\Sigma_{\rm gas}$ is given
in units of $M_{\odot}$~pc$^{-2}$. A least-squares fit to the models
listed in Table~\ref{tab1} (both low $T$ and high $T$) gives 
$A = (1.4 \pm 0.4) \times 10^{-4}$ and $\alpha = 1.45 \pm 0.07$, agreeing with
the observations to within the errors. 

Note LT models tend to have slightly higher SF rates than equivalent HT
models. Thus, observations may be able to directly measure the effective sound 
speed (roughly equivalent to velocity dispersion) of the star-forming
gas in galactic disks and nuclei. More simulations will be needed to
demonstrate this quantitatively. 

Our chosen models do not populate the lowest and highest star
formation rates observed. Interacting galaxies can produce very unstable disks
and trigger vigorous starbursts \citep*[e.g.,][]{li04}. Quiescent normal
galaxies form stars at a rate below our mass resolution limit. 
Our most stable models indeed show no star formation in the first few
billion years.

\section{STAR FORMATION THRESHOLD}

\begin{figure}
\begin{center}
\includegraphics[width=2.1in]{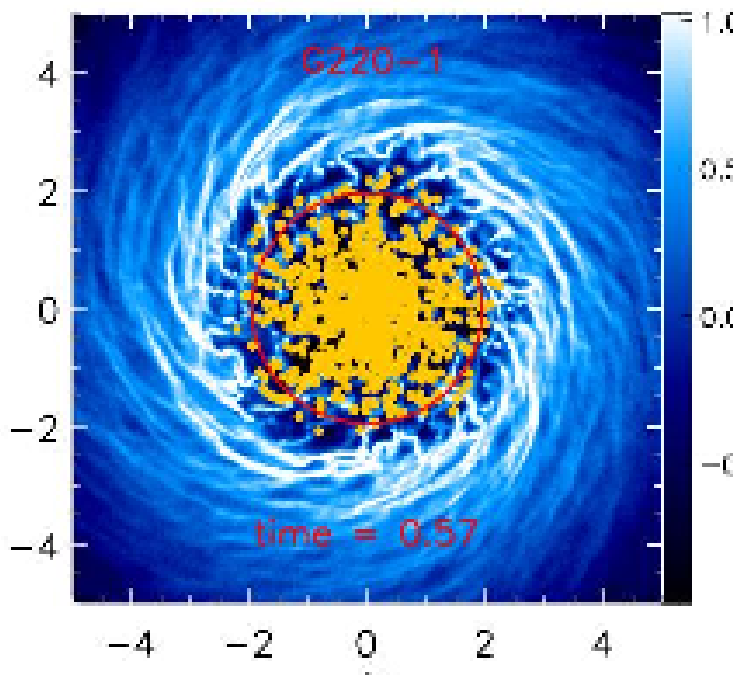}\\
\includegraphics[width=2.1in]{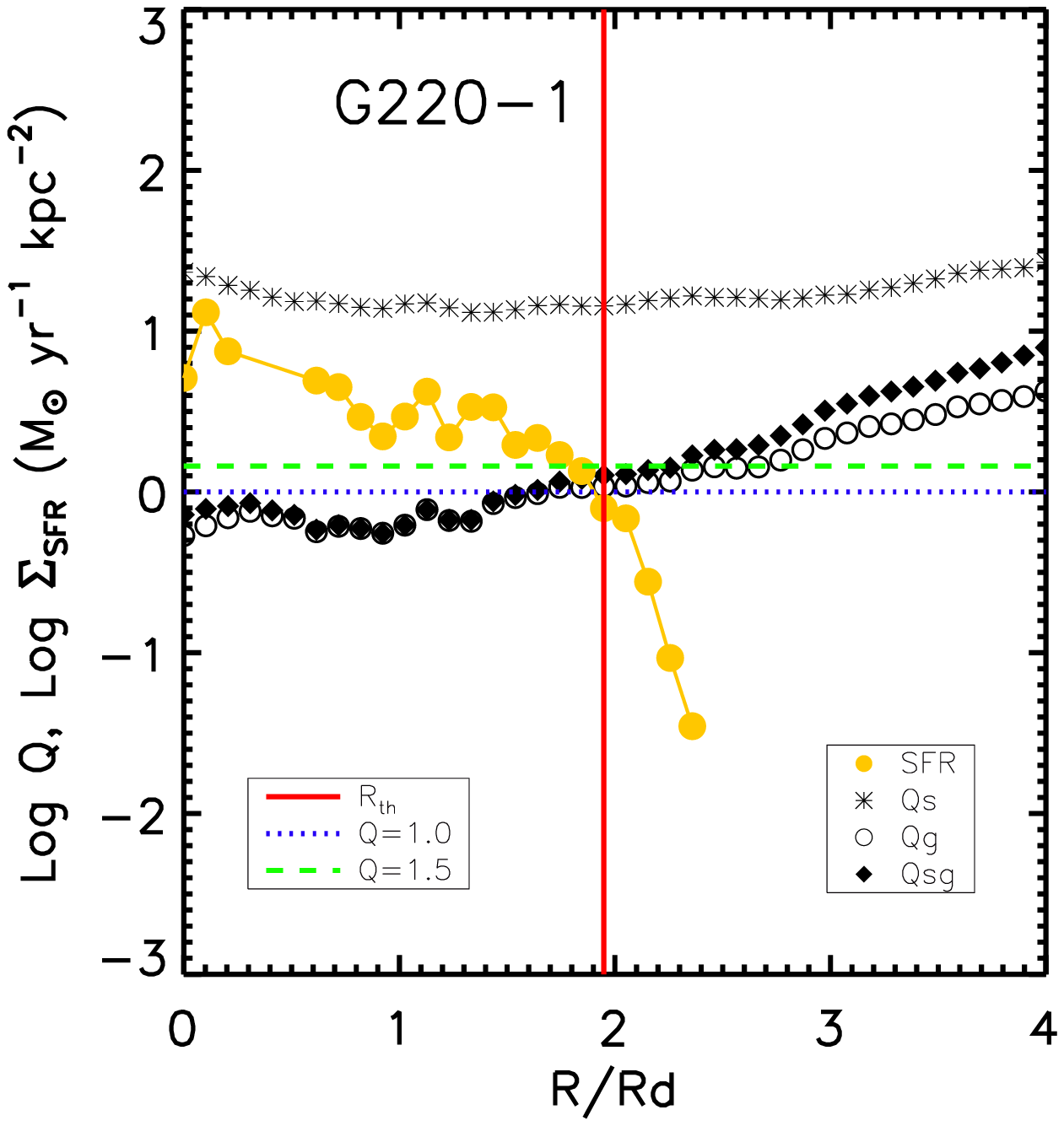}\\
\includegraphics[width=2.1in]{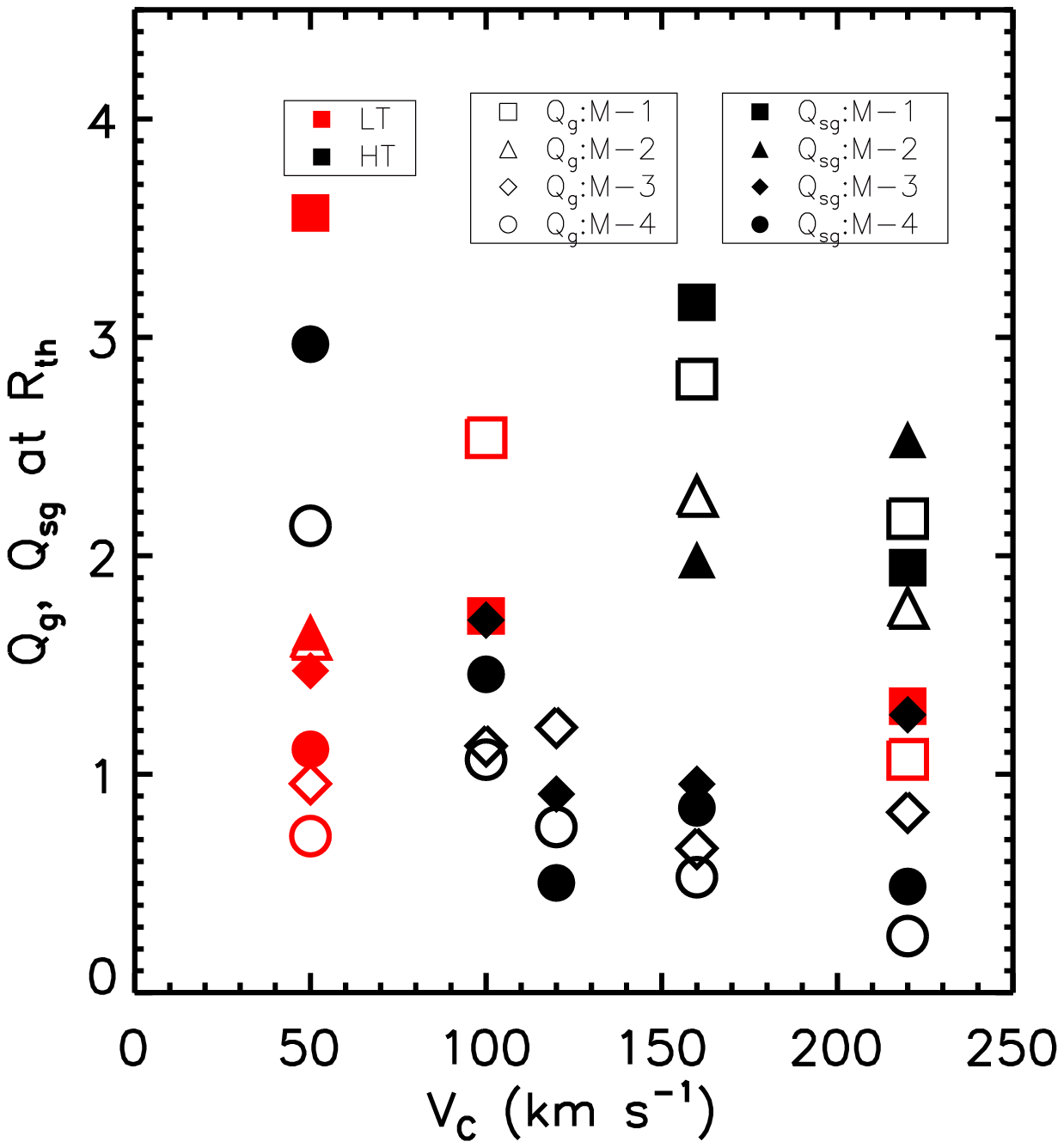}
\caption{\label{fig_threshold} {\em (Top)} Star formation threshold
  illustrated by the low $T$ model G220-1 with $N_{\rm tot} = 6.4 \times 10^6$. 
Log of gas surface density is shown,
  with values given by the color bar.  Yellow dots indicate SCs,
  while the red circle shows $R_{\rm th}$. {\em (Middle)} Radial
  profiles of star formation rate (yellow circles), and Toomre Q
  parameters for stars $Q_{\rm s}$ {\em (asterisks)}, gas $Q_{\rm g}$
  {\em (circles)}, and stars and gas combined $Q_{\rm sg}$ {\em
    (diamonds)}.  The red line shows $R_{\rm th}$.  \textit{Bottom:}
  critical values of $Q_{\rm sg}$ {\em (filled symbols)} and $Q_{\rm
    g}$ {\em (open symbols)} at $R_{\rm th}$ for both low {\em (red)} and
high {\em (black)} $T$ models. 
}
\end{center} 
\end{figure}

A threshold is clearly visible in the spatial distribution of gas and
stars in our galaxy models, as illustrated in Figure \ref{fig_threshold}. The
critical value of the instability parameter at threshold can be quantitatively
measured from the radial profile as indicated in the middle panel, which shows
a sharp drop of $\Sigma_{\rm SFR}$ at $R \sim 2R_{\rm d}$. The critical values
of $Q_{\rm sg}$ and $Q_{\rm g}$ at the threshold $R_{\rm th}$ are shown in the
bottom panel of Figure~\ref{fig_threshold} for all the fully resolved models
listed in Table~\ref{tab1}. 
The critical values of $Q_{\rm sg}$ appear to be generally higher than
$Q_{\rm g}$ in the same galaxy, and both have lower values ($<1$) in more
unstable models. 

Most galaxies not classified as starbursts have gas fractions
comparable to or less than our most stable models, so the observation
of a threshold value of $Q_{\rm g} \sim 1.4$ may reflect the stability
of the galaxies in the sample \citep{martin01}. Observed variations in
the threshold also appear to occur naturally. If we only use the
Toomre criterion for the gas $Q_{\rm g}$ we get slightly larger
scatter than if we include the stars and use the combined criterion
$Q_{\rm sg}$, but the effect is small.

\section{DISCUSSION AND SUMMARY}
\begin{figure}[h]
\begin{center}
\includegraphics[width=2.7in]{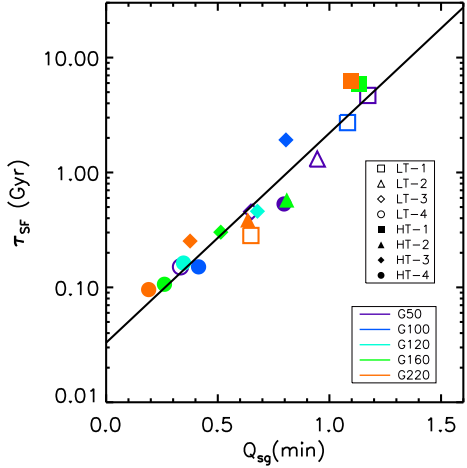}
\caption{\label{fig_tau} Star formation timescale $\tau_{\rm SF}$ as a
  function of initial $Q_{\rm sg}(min)$, for both low
  $T$ {\em (open symbols)} and high $T$ {\em (filled symbols)} models.  The
solid line is the least-square fit.} 
\end{center} 
\end{figure}

What controls star formation in different galaxies? Our models suggest the
answer is the nonlinear development of gravitational instability. Figure 
\ref{fig_tau} shows the correlation between the star formation
timescale $\tau_{\rm SF}$ and the initial minimum $Q_{\rm sg}(min)$
for fully resolved models listed in Table~\ref{tab1}.  The best fit is
$ \tau_{\rm SF} = (34 \pm 7\mbox{
  Myr})\times \exp \left[(4.2 \pm 0.3)Q_{\rm sg}(min)\right]$.
Quiescent star formation occurs where $Q_{\rm sg}$ is large, while
vigorous starbursts occur where $Q_{\rm sg}$ is small. 
This differs from the emphasis on supersonic turbulence by
\citet{kravtsov03}.  The maximum strength of instability $Q_{\rm
  sg}(min)$ depends on the mass of the galaxy and the gas fraction.
The larger the halo mass, or the larger the gas fraction, the smaller
resulting $Q_{\rm sg}(min)$, and thus the shorter $\tau_{\rm SF}$.

Typical observed starburst times of $10^8$ yr are consistent with our
fit for $\tau_{\rm SF}$ \citep{kennicutt98b}.  This 
also agrees with the observations by \citet{macarthur04} that the 
star formation rate depends on the galaxy potential.
\citet{mcgaugh00} show a break in the Tully-Fisher relation for
galaxies with $V_{\rm c} \le 90$ km s$^{-1}$, suggesting a transition
at this scale. Indeed, our models with $V_{\rm c} \le 100$ km s$^{-1}$
and gas fraction $\le 50\%$ of the disk mass appear to be rather
stable ($Q_{\rm sg} > 1.0$), with no star formation in the first 3
Gyrs, while models with $V_{\rm c} \ge 120$ km s$^{-1}$ become less
stable, forming stars easily. This is also consistent with
the rotational velocity above which dust lanes are observed 
to form \citep{dalcanton04}.

We have deferred inclusion of explicit feedback, magnetic
fields, and gas recycling to future work. However, we believe each will have
minor effects on the questions considered here. The assumption of an isothermal
equation of state for the gas implies substantial feedback to maintain
the effective temperature of the gas against radiative cooling and
turbulent dissipation. Real interstellar gas has a wide range of
temperatures. However, the rms velocity dispersion generally falls
within the range 6--12 km s$^{-1}$ (e.g., \citealt{elmegreen04}).
Direct feedback from starbursts may play only a minor role in quenching
subsequent star formation \citep[e.g.,][]{kravtsov03,monaco04}, perhaps
because most energy is deposited not in the disk but  
above it as superbubbles blow out \citep[e.g.,][]{fujita03,avillez04}. 
\citet{kim01} demonstrate that swing and magneto-Jeans instabilities operating
in a gaseous disk occur at $Q_{\rm g} \sim 1.4$, suggesting that magnetostatic
support is unimportant.  The lack of gas recycling both from disrupted
molecular clouds and from massive stars will change the detailed
patterns of star formation, but probably not the overall results.

Simulations of isolated, isothermal disks by \citet{robertson04} show
large-scale collapse in their centers leading to disks far smaller than
observed, which they argued was caused by an isothermal equation of
state. This behavior does not occur in our model with physical parameters
close to theirs, but resolution sufficient to resolve the Jeans length.
Similarly, \citet{governato04} argue that several long-standing
problems in galaxy simulations such as the angular momentum
catastrophe may well be caused by inadequate resolution, or violation
of the other numerical criteria. We will present more resolution studies in
future work. 

In summary, our models reproduce quantitatively not only the Schmidt law, but
also the star formation threshold in disk galaxies. We find a direct
correlation between the star formation rate and the strength of gravitational
instability. This suggests that gravitational instability in effectively
isothermal gas may be the dominant physical mechanism that controls
the rate and location of star formation in galaxies. 
Unstable galaxies were more common at early cosmic times, 
so our results, together with merger-induced starbursts \citep{li04}
may account for the Butcher-Oemler effect \citep{butcher84} of increasing
blueness of galaxies with redshift. Massive galaxies form stars
quickly, which may account for the downsizing effect that star
formation first occurs in big galaxies at high redshift, while modern
starburst galaxies are small (\citealt{cowie96, poggianti04, ferreras04}). The
slow evolution of star formation in our low mass models resembles that
observed in low surface brightness galaxies \citep{vandenhoek00}.

\acknowledgments We thank V. Springel for making both GADGET and 
his galaxy initial condition 
generator available, as well as for useful
discussions, A.-K. Jappsen for participating in the implementation of
sink particles in GADGET, and F. Adams, J. Dalcanton, R. Kennicutt, J.
Lee, C. Martin, D. McCray, T. Quinn, M. Shara, and J. van Gorkom for
useful discussions. The referee, F. Governato, also gave valuable
comments. This work was supported by NSF grants AST99-85392 and AST03-07854,  
NASA grant NAG5-13028, and DFG Emmy Noether grant KL1358/1.  Computations were
performed at the Pittsburgh Supercomputer Center supported by the NSF, on the
Parallel Computing Facility of the AMNH, and on an Ultrasparc III cluster
generously donated by Sun Microsystems.

\end{document}